\documentclass[fleqn,usenatbib]{mnras}

\usepackage{newtxtext,newtxmath}

\usepackage[T1]{fontenc}
\usepackage{ae,aecompl}

\usepackage{graphicx}	
\usepackage{amsmath}	
\usepackage{amssymb}	

\title[Temperature stability in LISA Pathfinder]{Temperature stability in the sub-milliHertz band \\ with LISA Pathfinder}

\author[Armano et al.]{
M~Armano$^{a}$,
H~Audley$^{b}$,
J~Baird$^{c}$,
P~Binetruy$^{c}$\thanks{Deceased 2017 March 30},
M~Born$^{b}$,
D~Bortoluzzi$^{d}$,
E~Castelli$^{e}$, \newauthor
A~Cavalleri$^{f}$,
A~Cesarini$^{g}$,
A\,M~Cruise$^{h}$,
K~Danzmann$^{b}$,
M~de Deus Silva$^{i}$,
I~Diepholz$^{b}$, \newauthor
G~Dixon$^{h}$, 
R~Dolesi$^{e}$, 
L~Ferraioli$^{j}$,
V~Ferroni$^{e}$,
E\,D~Fitzsimons$^{k}$,
M~Freschi$^{i}$,
L~Gesa$^{l}$, \newauthor
F~Gibert$^{e}$, 
D~Giardini$^{j}$,
R~Giusteri$^{e}$,
C~Grimani$^{g}$,
J~Grzymisch$^{a}$,
I~Harrison$^{m}$,
G~Heinzel$^{b}$, \newauthor
M~Hewitson$^{b}$, 
D~Hollington$^{n}$,
D~Hoyland$^{h}$,
M~Hueller$^{e}$, 
H~Inchausp\'e$^{c}$ $^{o}$, 
O~Jennrich$^{a}$, \newauthor
P~Jetzer$^{p}$, 
N~Karnesis$^{c}$,
B~Kaune$^{b}$,
N~Korsakova$^{q}$,
C\,J~Killow$^{q}$,
J\,A~Lobo$^{l}$\thanks{Deceased 2012 September 30},
I~Lloro$^{l}$, \newauthor
L~Liu$^{e}$, 
J\,P~L\'opez-Zaragoza$^{l}$,
R~Maarschalkerweerd$^{m}$,
D~Mance$^{j}$,
C~Mansanet$^{l}$,
V~Mart\'{i}n$^{l}$, \newauthor
L~Martin-Polo$^{i}$, 
J~Martino$^{c}$,
F~Martin-Porqueras$^{i}$,
I~Mateos$^{l}$,
P\,W~McNamara$^{a}$, \newauthor
J~Mendes$^{m}$, 
L~Mendes$^{i}$,
N~Meshksar$^{j}$,
M~Nofrarias$^{l}$,
S~Paczkowski$^{b}$,
M~Perreur-Lloyd$^{q}$, \newauthor
A~Petiteau$^{c}$,
P~Pivato$^{e}$,  
E~Plagnol$^{c}$,
J~Ramos-Castro$^{r}$, 
J~Reiche$^{b}$,
D\,I~Robertson$^{q}$, \newauthor
F~Rivas$^{l}$,
G~Russano$^{e}$,  
J~Sanju\'an$^{s}$,
J~Slutsky$^{t}$,
C\,F~Sopuerta$^{l}$, 
T~Sumner$^{n}$,
D~Texier$^{i}$, \newauthor
J\,I~Thorpe$^{t}$,
C~Trenkel$^{u}$,
D~Vetrugno$^{e}$, 
S~Vitale$^{e}$,
G~Wanner$^{b}$,
H~Ward$^{q}$, 
P\,J~Wass$^{n}$, \newauthor
D~Wealthy$^{u}$,
W\,J~Weber$^{e}$, 
L~Wissel$^{b}$,
A~Wittchen$^{b}$ and 
P~Zweifel$^{j}$
\\
\\
Affiliations are listed at the end of the paper
}

\pubyear{2019}

\begin{document}
\label{firstpage}
\pagerange{\pageref{firstpage}--\pageref{lastpage}}
\maketitle

\newpage 

\begin{abstract}
LISA Pathfinder (LPF) was a technology pioneering mission designed 
to test key technologies required for gravitational wave detection in space.
In the low frequency regime (milli-Hertz and below), where space-based gravitational 
wave observatories will operate, temperature fluctuations play a crucial role
since they can couple into the interferometric measurement and the test masses' free-fall 
accuracy in many ways. A dedicated temperature measurement subsystem, with noise 
levels in 10\,$\mu$K\,Hz$^{-1/2}$ down to 1\,mHz was part of the diagnostics unit on board LPF.
In this paper we report on the temperature measurements throughout mission operations,  
characterize the thermal environment, estimate transfer functions between different 
locations and report temperature stability (and its time evolution) at 
frequencies as low as 10\,$\mu$Hz, where typically values around 1\,K\,Hz$^{-1/2}$ were measured.
\end{abstract}




\section{Introduction}
\label{sec.intro}

LISA Pathfinder (LPF)~\citep{Anza05, Antonucci12} was an ESA mission with NASA contributions designed 
to test key technologies for the future gravitational waves observatory in space, 
the Laser Interferometry Space Antenna (LISA)~\citep{Amaro01}. 
LISA Pathfinder was  launched on 2015 December 3 
and started scientific operations at the Lagrange point L1 on 2016 March. 
The original mission plan included a six month operation period split between 
the two experiments on-board: the European LISA Technology Package (LTP) 
and the NASA Disturbance Reduction System (DRS). After an extended 
operations phase, the mission was finally decommissioned and
passivated on 2017 July. 

The main scientific goal of the mission was expressed in terms of a differential 
acceleration noise between two test masses 
in geodesic motion, i.e. in nominal free fall inside the spacecraft.  
LISA Pathfinder achieved residual acceleration noise levels
of $\rm (1.74 \pm 0.01)\,fm\,s^{−2}/\sqrt{Hz}$ above 2\,mHz, 
and $\rm (60\pm10)\,fm\,s^{−2}/\sqrt{Hz}$ at 20\,$\rm \mu$Hz~\citep{Armano16, Armano18}. 
Demonstrating this purity of free fall at these low frequencies was well
beyond the capabilities of any ground-based experiment.

Enabling a new observing window in the gravitational sky requires 
facing new technological challenges. 
The coupling of 
 low frequency temperature perturbations to instrument performance is one of these challenges. 
Indeed, temperature fluctuations will play an important role in space-borne gravitational 
detectors since typically their most significant contribution is in the low frequency (sub-milliHertz) part of the measurement window
where temperature driven effects may even limit the overall instrument sensitivity.
Moreover, at such time scales ---temperature changes that can last for hours--- 
they are ubiquitous to the satellite with a potential to impact different 
stages and subsystems of the measuring chain, e.g. thermal induced forces~\citep{Carbone07}
acting on the Gravitational Reference Sensor (GRS)~\citep{Dolesi03}, temperature induced path-length 
variations~\citep{Nofrarias07,  Nofrarias13, Gibert15} in the optical metrology subsystem (OMS)~\citep{Heinzel04} and
thermal effects in all associated electronics.

Unlike most space missions, where the satellite house-keeping system is in charge 
of monitoring the environment, LISA Pathfinder included a precision diagnostics subsystem~\citep{Canizares09} 
designed with a two-fold objective. First, to monitor noise disturbances 
and, second, to study the contribution of these disturbances to the instrument noise budget. 
The diagnostic subsystem was composed of sensors (magnetometers, temperature
sensors and a particle counter) and actuators (heaters and coils). The latter were used 
to induce controlled perturbations, which allowed us to derive coupling factors and transfer functions between 
thermal and magnetic perturbations and the outputs of the GRS and the OMS.

In this paper we focus on the thermal environment during the LISA Pathfinder mission and 
report on the achieved on-board temperature stability.
During periods of uninterrupted operation we see a temperature 
stability of tens of microKelvin in the milliHertz measuring band.
We also identify the most significant drivers of thermal disturbances at milliHertz
frequencies and suggest how these should be addressed in a future LISA mission.
While the reported experiments were designed to answer questions for the LISA mission 
there is an increasing demand for very controlled and stable environments
in a wide range of experiments, both on ground and in space and these results will have a wider applicability. 
Examples include geodesy missions~\citep{Sheard12}, on-going or proposed fundamental 
physics~\citep{Aguilera14, Lammerzahl01} missions
to on ground experiments aiming at exoplanets detection~\citep{Stefansson16}. 
In all these cases temperature 
stability is crucial to suppress spurious effects arising from a large variety of thermal coupling phenomena.
 
The manuscript is organized as follows: in Section~\ref{sec.diagnostics} we describe the temperature diagnostics items, their role 
and distribution in the instrument. Section~\ref{sec.temp_profiles} is devoted to the description of the temperature evolution during
the mission timeline as measured by the diagnostics subsystem while Section~\ref{sec.spectra} gives a 
more detailed insight in the noise performance and derives temperature couplings ---transfer functions--- 
between different sensitive locations in the instrument. We end with a final discussion where we provide our conclusion 
and implications for LISA.

\section{The thermal diagnostic subsystem}
\label{sec.diagnostics}
 
The temperature diagnostic subsystem on board LISA Pathfinder was designed primarily 
to monitor temperature fluctuations in sensitive locations on board the satellite. 
This section describes the elements composing the temperature diagnostics subsystem 
and their distribution in the satellite.  There are 24 thermal sensors and 
18 heaters distributed around the LTP Core Assembly~\citep{Canizares09} as we show in Figure~\ref{fig.Location_heaters_sensors}. 
In the following section we summarise the main characteristics that describe this subsystem.  

\subsection{Sensors distribution and rationale }

\begin{figure}
\includegraphics[width=0.5\textwidth]{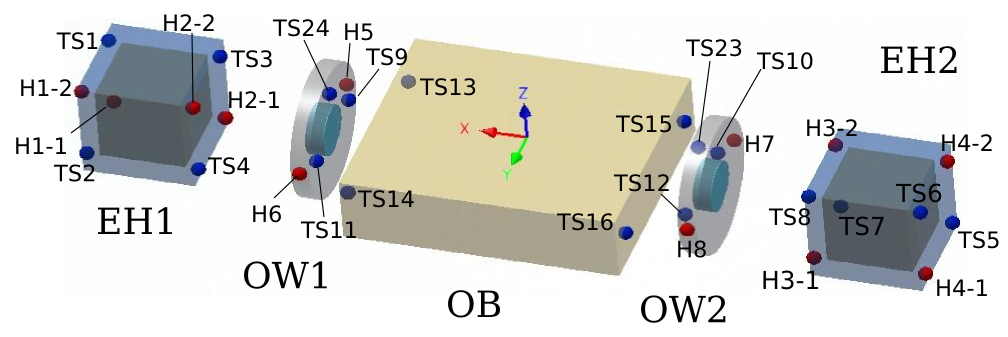}
\includegraphics[width=0.5\textwidth]{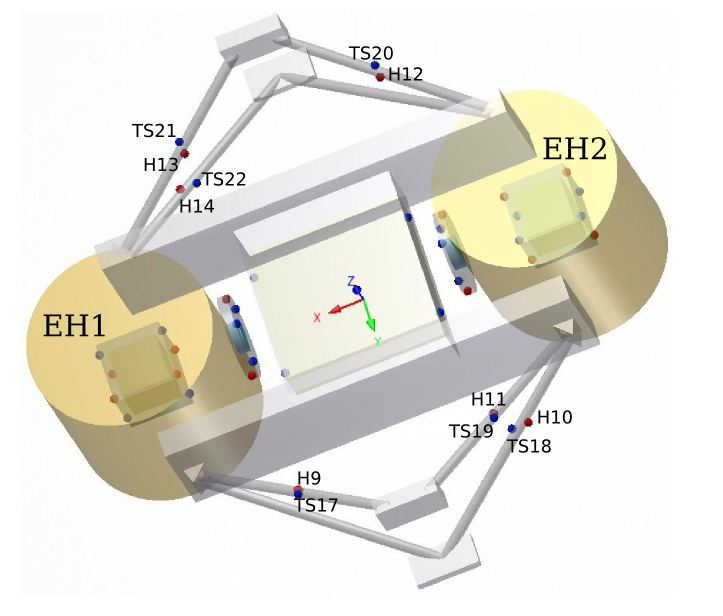}
\includegraphics[width=0.5\textwidth]{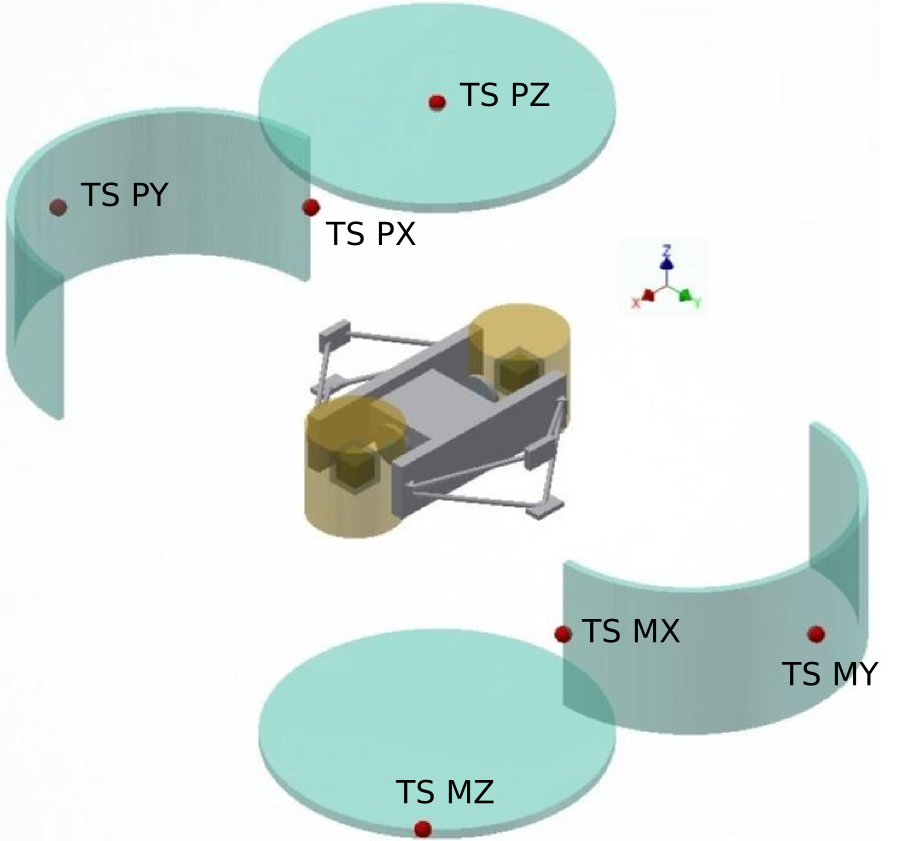}
\caption{Distribution of temperature sensors in LISA Pathfinder. 
\emph{Top:} Heaters and temperature sensors in the electrode housing (EH), the optical window (OW) and the optical 
bench (OB) from the diagnostics subsystems. \emph{Middle:} Temperature sensors and heaters in the struts from the diagnostics subsystems. \emph{Bottom:}
Some of the platform sensors attached to thermal shield surrounding the instrument.
}
\label{fig.Location_heaters_sensors} 
\end{figure}

Temperature sensors were distributed across the instrument in order to monitor critical locations where 
temperature fluctuations could perturb the main scientific measurement on-board, the 
differential acceleration between the two free falling test masses. 
The distribution of temperature sensors in the instrument ---see Figure~\ref{fig.Location_heaters_sensors}--- 
was chosen to study the effects that can produce such perturbations. 
These were basically divided in two families: i) thermal induced forces directly 
applied on the test masses and ii) thermo-elastic distortions  
of the optical system.

The first kind of effect is highly dependent on the gradient of
temperature across theGRS housing, hence two temperature sensors
were attached on both sides of the sensitive axis (the axis that joins
both test masses) of the electrode housing (EH) containing the
test masses. The main instrument, consisting of both the test masses
inside the vacuum enclosure and the optical bench (OB), was hosted
inside a thermal shield to protect it from the fluctuations associated
with the electronic boxes. The struts holding the instrument inside
the thermal shield were the thermal link to the outer environment
and therefore temperature was monitored in six of them, from a
total of eight. Three sensors were attached to each of the two
optical windows (OWs), the optical elements enabling the laser
link between the OB and the test masses. Finally, four temperature
sensors were located on the OB, one on each corner, to monitor
gradients which could potentially induce thermoelastic induced
bending.

The satellite also included 
sensors monitoring temperature for each of the different units in the platform. 
These were located outside the thermal shield containing the instrument and,
were primarily for monitoring unit health so had 
less stringent requirement in terms of precision or stability than the sensors of the temperature diagnostics subsystem.
They are however indicative of the temperature fluctuations of the electronic units and,
hence, useful as a reference of the temperature environment surrounding our instrument. 

\subsection{Electronics and sensor performance}
\label{sec.diagnostics_electronics}

On-ground estimates of the effects which can potentially disturb 
the main free fall measurement set a requirement for temperature stability of 
$100\rm\, \mu K /\sqrt{Hz}$ in the LISA Pathfinder measuring bandwidth, $1\,{\rm mHz} \leq f \leq 30\,{\rm mHz}$.
In the design phase of the instrument it was decided that the temperature measurement subsystem 
had to be able to clearly distinguish such levels of disturbances and therefore the thermal sensitivity requirement was set to $10\rm\, \mu K /\sqrt{Hz}$ in the same measurement band~\citep{Lobo06a}. 

The sensors selected to achieve this goal were Negative Temperature Coefficient (NTC) 
thermistors (Betatherm G10K4D372) with a nominal resistance of 
10 k$\Omega$. These were considered to be the best candidate to achieve the demanding 
requirements~\citep{Sanjuan07}. 
On the downside, the oxides used to manufacture thermistors contain magnetically active materials.
This can have a potential impact when used in locations close to the test mass
as they can introduce a source of local magnetic gradient.
The studies performed showed, however, that the potential impact of this effect 
was substantially reduced by means of demagnetisation procedures and hence 
was estimated to not impact the instrument performance~\citep{Sanjuan08_mag}.

The front-end electronics was based on an AC powered Wheatstone bridge~\citep{Sanjuan07}.
A constraint to be taken into account in the design of the subsystem was the dissipated power 
in the temperature sensors: it was limited to 10\,$\rm \mu W$ in order to prevent both, thermal effects 
appearing in critical subsystems to which
the sensors were attached, and to avoid self-heating errors in the sensor.  
In order to improve the resolution, six scales were defined with center temperatures 
12, 15, 20, 22.5, 
25, and 27.5$^{\circ}$\,C which, at the same time, 
kept the output voltage of the bridge close to zero. 
Change in temperature scales produced  
spikes in the read-out as a consequence of the different zero in each temperature scale. 
These spikes only appeared during phases of high temperature drifts, for example
temperature experiments or changes in the spacecraft configuration,  
and never during phases of scientific runs where the instrument was kept unperturbed to achieve the highest degree 
of free fall. The spikes were suppressed from the time series by means of data analysis post-processing. 


As commented above, the temperature diagnostics subsystem included heaters with the aim 
of injecting controlled temperature perturbations in sensitive locations and study their coupling to
instrument performance.
Two different types of heaters were used: Kapton heaters (45 $\Omega$ resistors) with
a maximum power of 2\,W were attached to the lateral sides of the OWs and in the struts whilst 
the heaters inside the EH were thermistors (2 k$\Omega$ resistors) 
acting as heaters, with a maximum power of 45\,mW. The decision to use the latter was driven by  the stringent 
contamination requirements inside the EH. 

\section{Temperature evolution during operations} 
\label{sec.temp_profiles}

\begin{table}
       \caption{Dates associated with events that impacted the thermal balance onboard LISA Pathfinder. 
       In parenthesis we include the Days After Launch (DAL).  \label{tab.key_dates_events} }
   \begin{tabular}{lc}
   \hline
          Event  &  Date (DAL) \\
         \hline 
          (\textcolor{red}{a}) Propulsion module released   		& {22 Jan '16 (50)}   \\
          (\textcolor{red}{b}) TMs released  								& {15 Feb '16 (75)}  \\
          (\textcolor{red}{c}) DMU SW crash 							& {05 May '16 (154)}   \\
          (\textcolor{red}{d}) Cluster-2 DCIU anomaly 			& {09 Jul '16 (219)}  \\
          (\textcolor{red}{e}) LTP safe mode   							& {24 Sept '16 (296)}   \\
          (\textcolor{red}{f}) DMU SW crash and reboot 			& {21 Oct '16 (323)} \\
          (\textcolor{red}{g}) Thruster-4 anomaly					& {27 Oct '16 (329)} \\
          (\textcolor{red}{h}) TMs grabbed and TMs released 	& {15 Jan '17 (409)} \\
          (\textcolor{red}{i}) Cooling down 								& {23 Jan '17 (417)}\\
          (\textcolor{red}{j}) Cooling down 								& {29 Apr '17 (513)}\\
          (\textcolor{red}{k}) Switch of SAU 								& {02 Jul '17 (577)}
    \end{tabular}
\end{table}

\begin{table}
      \caption{Dates associated with thermal experiments onboard LISA Pathfinder. 
      In parenthesis we include the Days After Launch (DAL).  \label{tab.key_dates_experiments} }
    \begin{tabular}{|cc|}
   \hline
  Event  &  Date  (DAL)  \\
      \hline

(1) Thermal injections in EH1 & {10 Mar '16 (98) }   \\
(2) Thermal injections in EH1 & {28 Mar '16 (116)}  \\
(3) Thermal injections in EH2 & {28 Mar '16 (116)}  \\
(4) Thermal injections in EH1 & {17 Apr '16 (136)}  \\
(5) Thermal injections in EH2 & {18 Apr '16 (137)}   \\
(6) Thermal injections in EH1 & {25 May '16 (174)} \\
(7) Thermal injections in EH2 & {27 May '16 (176)} \\
(8) Thermal injections in OWs & {13 Jun '16 (193)} \\
(9) Thermal injections in STRs & {13 Jun '16 (193)} \\
(10) Thermal injections in EH1 & {14 Nov '16 (347)} \\
(11) Thermal injections in EH2 & {15 Nov '16 (348)} \\
(12) Thermal injections in EH1 & {18 Jan '17 (412)} \\
(13) Thermal injections in EH2 & {19 Jan '17 (413)} \\
(14) Thermal injections in OWs & {17 Jun '17 (562)} \\
(15) Thermal injections in EH1 & {24 Jun '17 (569)} \\
    \end{tabular} 
 \end{table}
	
\begin{figure*}
\centering
\includegraphics[width=1\textwidth]{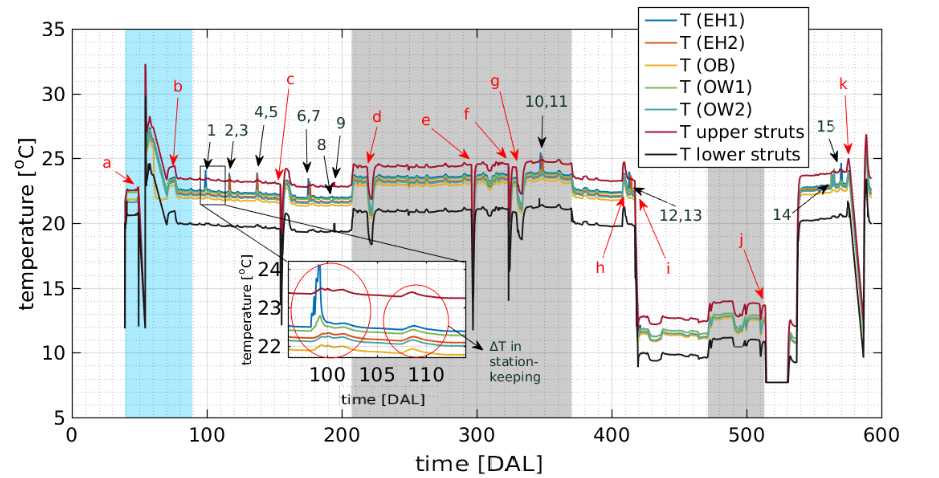}
\includegraphics[width=1\textwidth]{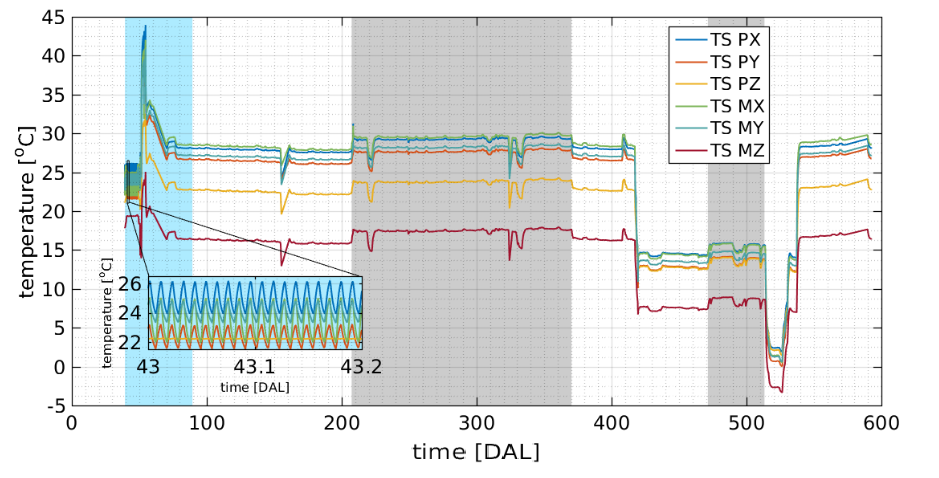}
\caption{Temperature evolution during the whole mission time-line. 
The time axis  is  days after launch (DAL).
The initial 
cyan area (DAL 40-90) corresponds to the commissioning. The two grey shaded areas (DAL 210-370 and 470-510) correspond 
to the DRS operations, and the rest are LTP operations. \emph{Top:}  Temperature as measured by 
the diagnostics subsystem located 
in sensitive location of the LTP instruments, namely the optical window (OW),
the optical bench (OB), the electrode housing (EH) and the struts holding the 
LTP inside the thermal shield. The traces show the average temperature in locations with more than one sensor
(4 in the electrode housing and 3 in the optical window)
\emph{Bottom:} Temperature evolution during the whole mission time-line as measured by 
the platform sensors attached to the outer face of the thermal shield surrounding the LTP.
\label{TS}}
\end{figure*}

LISA Pathfinder was launched from the French Guiana 
on 2015 December 3. 
It took approximately a month --including LEOP (Launch and Early Orbit/Operations Phase) and apogee increase manoeuvres-- 
to reach the L1 orbit and start commissioning phase. LTP commissioning started on January 11th 
2016 and lasted until March 1st, when the mission started operations phase. 
LISA Pathfinder underwent different phases during scientific operations 
which included nominal operations for the two experiments on-board, the LTP
and the DRS, and an extended period of operations for both of them. 
Figure~\ref{TS} shows the temperature as measured during all the mission by 
the diagnostics subsystem thermistors located on the LTP (top panel) and 
by the platform sensors attached to the external face of the thermal shield surrounding the
LTP instrument (bottom panel). In the current section we provide some insight on the operations timeline 
from the point of view of the temperature evolution in the satellite as
provided by the previous sensors.

\subsection{Commissioning}

The commissioning was a particularly active period in terms of temperature variations.  
It started during the first days of January 2016, 
after a cruise phase to L1. 
The first set of operations relevant from the temperature point of view was the
\emph{bang-bang} temperature control active during the first days of operations. 
As seen in the inset of the Figure~\ref{TS} lower panel, it produced $\sim 2\,^{\circ}$C 
peak-to-peak variations with a frequency around 1\,mHz
as measured by the platform sensors
which was also measured, after the thermal shield attenuation, by the higher precision 
diagnostics sensors inside. 
As we will show in Section~\ref{subsec.transfer_functions} this allows an experimental determination
of the thermal shield thermal transfer function.

After this first phase and once in the desired orbit around L1, the propulsion module was ejected. 
This corresponds to the data gap in the commissioning phase 
---see the marked event in Figure~\ref{TS}--- 
since the LTP instrument was switched off during this action. 
The sudden temperature increase afterwards is an intended heating of the instrument 
to increase the out-gassing rate and improve the cleanliness conditions before the release of the 
test mass and the start of free fall operations. The release of the test masses took place 
on February 15th and 16th and it appears in the temperature read-out as a temperature 
increase due to associated satellite operations. 

\subsection{LTP operations}

The two periods of LTP operations took place from 2016 March 1
to 2016 June 26 in its nominal phase and from 2016 December 8
to 2017 March 17 and 2017 May 1 to 2017 June 30 in its extended
phase. 
In Figure~\ref{TS} we show the temperature evolution as measured by the temperature diagnostics subsystem.
A first characteristic to notice from this is a constant $3.5\,^{\rm o}$C gradient between the upper and lower struts during the 
whole mission, which only goes below the  $3\,^{\rm o}$C  during the cool down phase that we explain below. 
A much lower temperature difference is also observed between the temperature sensors 
located in the rest of LTP locations. In analysing these, it must be taken into account that the temperature sensors 
of the LTP diagnostics subsystem were optimised for precision and not for accuracy,
showing typically an absolute uncertainty $\sim 0.2\,^{\rm o}$C.
Therefore the differences in absolute temperature below this value must be considered 
within the error of the temperature read-out.  

The temperature evolution during the LTP operations phase shows a series of sudden decreases of 
temperature together with some other, smaller, temperature increases. 
In Table~\ref{tab.key_dates_events} we gather the main features 
that caused the temperature decreases in the time-line.
Most of these correspond to pauses in normal satellite operations, for example
anomalies of the thrusters subsystem during DRS operations or reboots of the Data Management 
Unit (DMU) --- the computer of the LTP instrument --- in the LTP phase operations. These events 
can trigger the satellite safe mode which switches off of some electronic units onboard, causing a 
consequent change in the satellite thermal balance.

The period of operations at lower temperature starting 430 days after launch corresponds to an intended 
cooling down of the spacecraft with the objective of study the instrument performance at a different temperature working point.
The series of measurements that took place during these weeks could successfully determine a decrease in the instrument acceleration
floor noise due to a suppression of the Brownian noise contribution due to gas particles hitting the test masses~\citep{Armano18}.
There were two cooling down phases ---see the dates in Table~\ref{tab.key_dates_events}. The first one decreased the temperature 
$\sim 10\,^{\rm o}$C leaving the housing surrounding the test masses at $\sim 12\,^{\rm o}$C. 
The second cooling down went below the design range of the temperature diagnostics subsystem and therefore 
appears as a saturated line in the temperature read-out in the upper panel  of  Figure~\ref{TS}. The lower panel in the 
same figure shows however the temperature in the thermal shield and how the coolest sensor in this structure reached 
$\sim -3\,^{\rm o}$C. Given the complete time series and from the surrounding temperatures, we can estimate a temperature 
in the test mass electrode housing in the housing for this period $\sim 2\,^{\rm o}$C. 

The LPF operations 
time line was planned to get the maximum scientific yield from its capabilities as technology demonstrator.
As such, an important fraction of the operations time was dedicated to experiments to gain insight in the 
different mechanism that can perturb the free fall. Temperature fluctuations were a relevant part of the noise 
budget and, hence, several experiments were planned during the mission duration. Table~\ref{tab.key_dates_experiments} 
shows the list of executed experiments in the different locations where the instrument included 
heaters from the diagnostics subsystem, as we have previously described in Section~\ref{sec.diagnostics_electronics}.
These experiments ---with temperature increases $\sim 2\,^{\rm o}$C and some of them lasting days--- 
had an important impact in the temperature profile and for that reason were carefully planned in advance. 
The typical impact is shown in the inset of the top panel of Figure~\ref{TS}. 
The increase in temperature when injecting thermal signals in one Electrode Housing (EH) is 
$\sim 2\,^{\rm o}$C for the sensors located on the same EH and $\sim 0.1\,^{\rm o}$C for the sensor on the other 
EH, $\sim 0.2\,^{\rm o}$C for the sensors on the optical bench, $\sim 0.4\,^{\rm o}$C for the sensors on the nearest OW 
and $\sim 0.15\,^{\rm o}$C on the farther OW. In comparison, heaters located on the optical 
windows and struts produced a temperature increase of $\sim 1\,^{\rm o}$C, 
as measured by temperature sensors on these locations.
 
For the particular case shown in Figure~\ref{TS}  
a series of temperature modulations were applied to the electrode housing producing 
a modulating force in the test mass that was used to derive the amount of coupling of the thermal induced forces in
the test mass motion. The modulation pattern was repeated at different absolute temperatures ---increasing 
the applied DC power-- producing a stair-like profile. The total experiment duration was about a day 
and it took LTP 2-3 days to recover from this disturbance.
 In Section~\ref{subsec.transfer_functions} we will come back to these injections to quantitatively 
characterise these thermal links.
			
The metastable orbit in L1 forced periodic station-keeping manoeuvres to keep the satellite in the predetermined 
orbit. Once the propulsion module was jettisoned ---during the commissioning phase--- the microNewton propulsion subsystem
was the only available thrusting system in LPF. 
The low thrust available required prolonged operations of the thrusters, initially set to one 12h period each week.  
In thermal terms, the result of this operation is shown again in the inset of the top panel of 
Figure~\ref{TS}, i.e. the operation of the thruster subsystem in wide range mode (with less precision but higher thrust) for 
station-keeping turned into a  $\sim 0.1\,^{\rm o}$C homogeneous increase in the different locations of the satellite.
Although not representing an operational inconvenience, this mode of operations resulted in a reduction of the mission effective 
duty cycle since the scientific runs required a extremely quiet environment.
Hence, the scientific runs duration were limited by the long time scales needed for the thermal environment to recover the steady state after station keeping. 
Later in the mission, this limitation was overcome by allowing longer, less frequent station keepings. 
That way, week-long scientific runs were executed which allowed the estimation of the instrument acceleration noise 
down to the 20\,$\rm \mu Hz$ regime~\citep{Armano18}. This is an important lesson for LISA as we point out in our final section. 
We can also see a thermal gradient between the temperatures measured by the sensors located on the LCA, with a maximum gradient of $\sim 13\,^{\rm o}$C, as shown Figure~\ref{TS} (bottom). During the first cooling down, the lower 
temperature given by the sensor TS MZ reaches a temperature of $\sim 7\,^{\rm o}$C, while during the second 
cooling down it reaches a negative value, $\sim -3\,^{\rm o}$C, the only temperature registered below zero.

\subsection{DRS operations}

DRS operations took place from 2016 June 27 to 2016 December
7 in its nominal phase and from 2017 March 18 to 2017 April 30
in its extended phase. 
The hand over to the DRS team required a series of configuration changes which had an impact on the
thermal environment. 
As thoroughly described in~\citep{Anderson18}, 
the hand over to the DRS team implied that, while the LTP instrument was still providing measurements 
of Spacecraft  and test masses attitude, this information was sent to the Integrated Avionics Unit (IAU),
which determined the forces and torques to be applied to the test masses and to the Spacecraft. 
The first were delivered again to the LTP while the second were sent to the two 
Colloidal MicroNewton Thrusters Assemblies (CMTAs). Both IAU and CMTAs were inactive during LTP operations. 

As seen in Figure~\ref{TS}, the continuous operations of these two units (gray shaded areas) changed the 
thermal balance in the satellite, raising the overall temperature by $\sim 2\,^{\rm o}$C with respect to
the LTP operations.
The DRS operations phase shows some pronounced decays in temperature 
corresponding with short interruptions of operations due to detected anomalies 
in the thruster subsystem. The reader is referred to \citep{Anderson18} for more detail on these.

\section{Temperature stability in LISA Pathfinder}
\label{sec.spectra}

\begin{figure}
\includegraphics[width=0.49\textwidth]{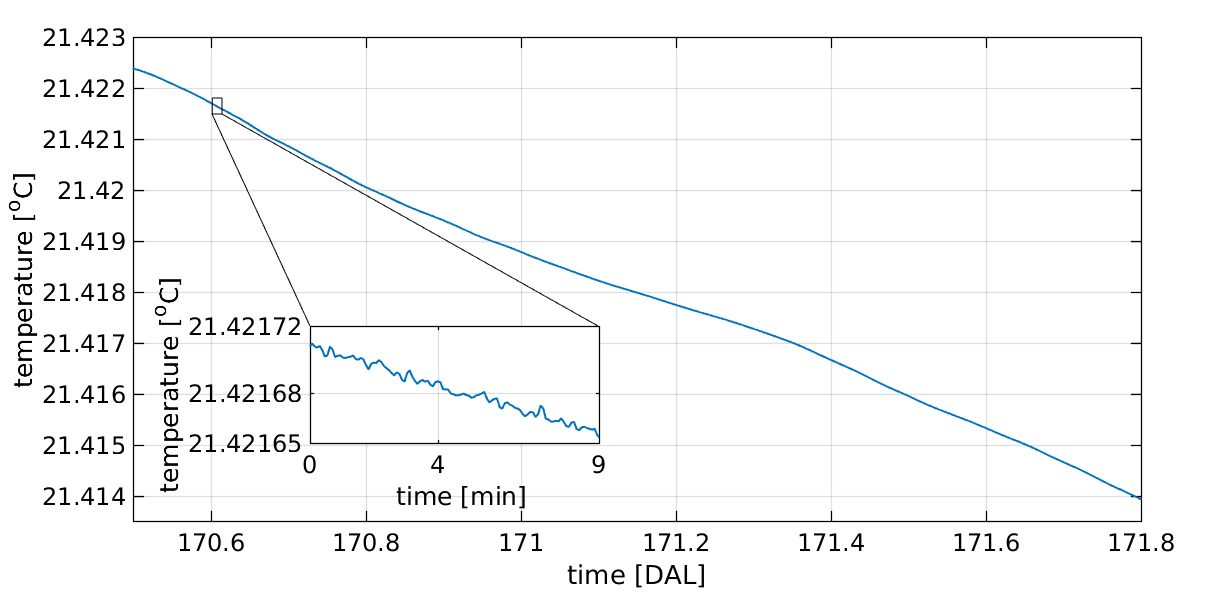}
\includegraphics[width=0.49\textwidth]{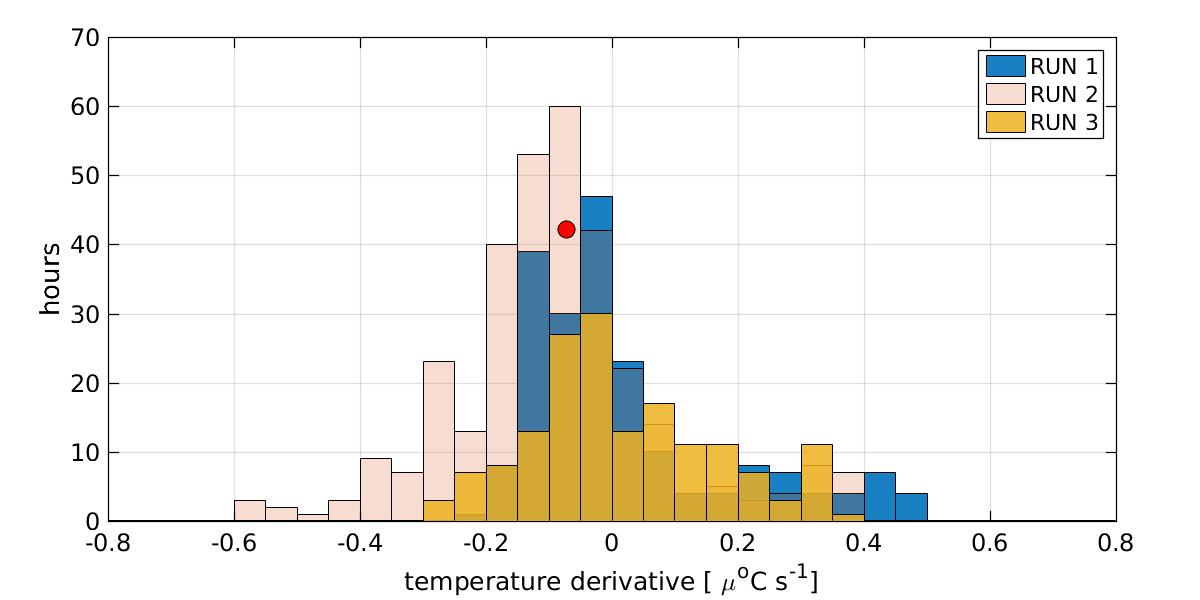}
\caption{\textit{Top}: Typical temperature evolution of a sensor located in the optical bench during a quiet noise run. 
\textit{Bottom}: Histogram showing the number of days with a given temperature drift for a sensor in the optical bench (TS13) for three different noise runs, with RUN 1 from November 17 to November 26 (2016), RUN 2 from February 14 to February 27 (2017) and RUN 3 from May 29 to June 5 (2017). The dot indicates the temperature derivative that corresponds to the time series at the top of the plot. 
\label{diff_TS13} }
\end{figure}

Achieving the scientific goal in LISA Pathfinder is a complex endeavour 
that requires excellent design and performance of several sensors and 
actuators. Temperature stability is a crucial one among them 
since it can impact in several ways the measurement chain, as 
we have previously described. The aim of the current 
section is to quantitatively assess the temperature 
stability of our instrument which, as we will 
see, goes in parallel with the determination 
of the performance of our temperature diagnostics
subsystem. 

In Figure~\ref{diff_TS13} we evaluate the typical stability measured in LPF during
a scientific run. We use a sensor in the OB to show the typical
evolution of the temperature during a quiet interval. To give a more
comprehensive view we include in the bottom panel of Fig. 3 an
histogram with the temperature derivative for a sensor in the OB
during three noise runs. Overall, the instrument spent nearly 250 days 
with temperature drifts in the range $\pm 1\:\mu ^{\rm o}{\rm C}\,{\rm s}^{-1}$. 
Temperature drifts are not only an important figure for the 
platform stability but, as we comment below, they have 
an important role in the evaluation of the performance of our sensors. 

\begin{figure*}
\centering
\includegraphics[width=0.8\textwidth]{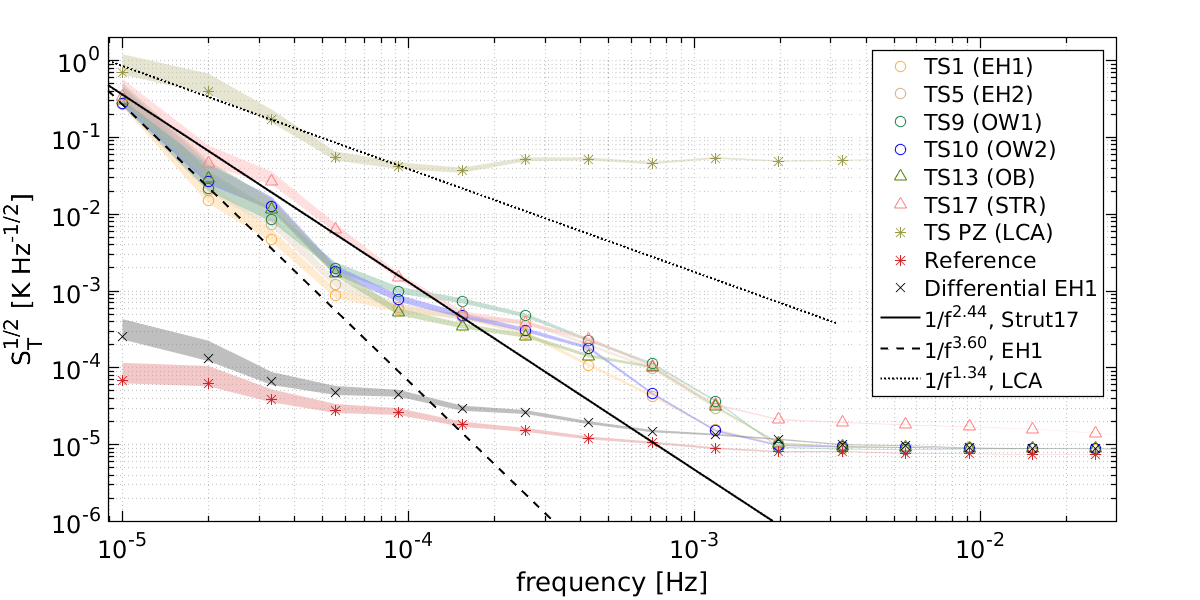}
\includegraphics[width=0.8\textwidth]{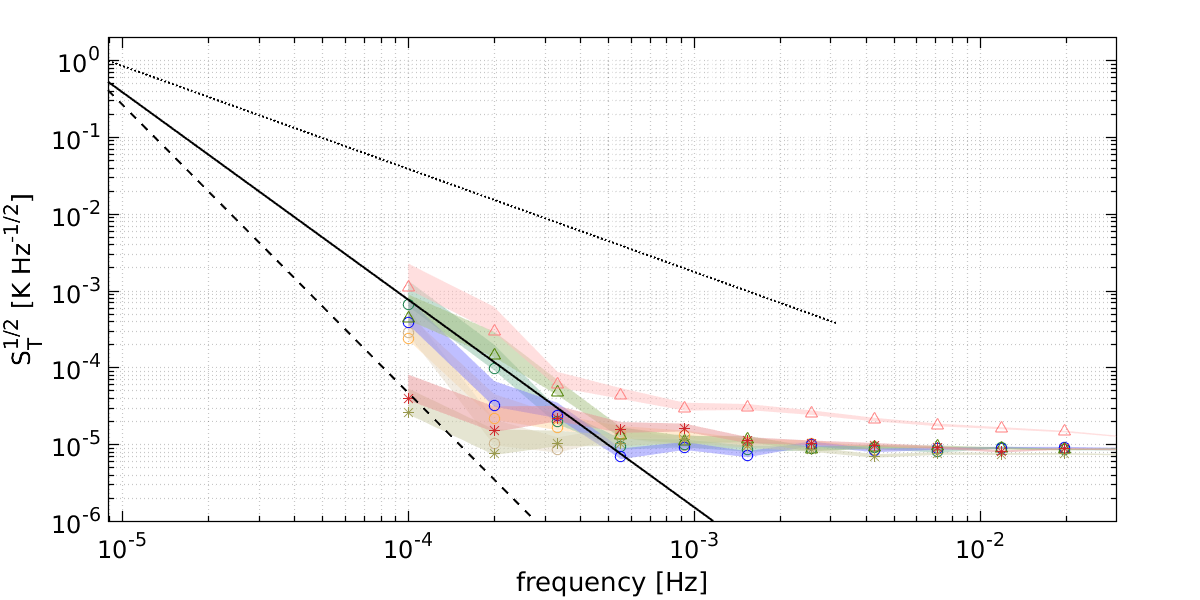}
\caption{Temperature stability measured as amplitude spectral density in different locations in LTP during the period 2017 February 14–27.  
\textit{Top}: Different locations inside LTP as measured by the temperature diagnostics subsystem. When different sensors were available we use the mean value of the measurements to obtain the fit at low frequencies. In addition, we show the temperature given by one space-craft temperature sensor in the outer face of the instrument thermal shield, also showing the fit for low frequencies taking into account the mean value of the measurements given by all the sensors located on the shield. We also show differential and reference measurements. 
\textit{Bottom}: The same as before evaluated in a shorter segment (2017 February 19–20) with lower temperature drift.  
 \label{fig.TS_psd}	}
\end{figure*}	

\subsection{Temperature fluctuations amplitude spectral density}
\label{subsec.power_spectral_density}

In order to evaluate the temperature stability in the instrument we start evaluating the stability 
of the environment surrounding the instrument. As previously described, the LTP 
was encapsulated in a thermal shield to suppress temperature fluctuations arising 
in the electronics surrounding the experiment. 
A series of sensors ---not belonging to the diagnostics subsystems---
monitored the temperature fluctuations at various locations around the satellite. 
These sensors were not designed for precise measurements 
of temperature stability, and therefore their 
noise floor, $\rm 50\,mK/\sqrt{Hz}$, is orders of magnitude above the ones belonging 
to the diagnostics subsystem.
Despite its reduced precision, the temperature fluctuations of the spacecraft are high 
enough at low frequencies to measure them. 

To evaluate temperature fluctuations we compute the amplitude spectral density 
---the square root of the power spectral density--- by means of the the Welch 
averaged periodogram. We use segments of 400 000\,s and apply Blackman-Harris window to prevent 
 spectral leakage. To make sure that the window is not biasing our estimate 
we get rid of the lowest four frequency bins of the spectra.
With the remaining spectra we evaluate the power 
at low frequency by means of a power law fit at the four lowest frequency bins. 
In the locations where we have more than one sensor we use an average of sensors when considering 
the power law fit. For the struts, where each sensor is attached to a different strut, 
we use the TS17 sensor as a typical case to evaluate the low frequency power.
For the interested reader we provide the coefficients of the fit 
obtained in each individual location in Appendix~\ref{sec.App_fits}.

In Figure~\ref{fig.TS_psd} we show our results for a long stable run on February 2017. 
In the top panel we can distinguish a thermally induced $f^{-1.34 \pm 0.05}$ 
power law below  $\rm 100\,\mu Hz$ in all the six sensors attached to the 
thermal shield. These low frequency fluctuation are 
transmitted to the instrument through the thermal shield and the struts.
As the spacecraft induced temperature fluctuations leak into the instrument they are 
successively suppressed, since each stage acts as a thermal low-pass filter. 
This can be appreciated when comparing the different slopes of the power law fits in Figure~\ref{fig.TS_psd}.
We notice how low frequency power has decreased from the original $f^{-1.34 \pm 0.05}$ 
on the outer layer of the thermal shield to $f^{-2.44 \pm 0.06}$ in the struts 
and even further to $f^{-3.60 \pm 0.04}$ if we continue to the EH sensors.
The cause of the decrease in power at lower frequencies is, as previously said,
the different layers of materials that the heat flow need to cross, which act 
as a series of consecutive thermal low-pass filters. 
In Section~\ref{sec.spectra} we will quantify and model this behaviour. 

\begin{figure*}
\includegraphics[width=0.8\textwidth]{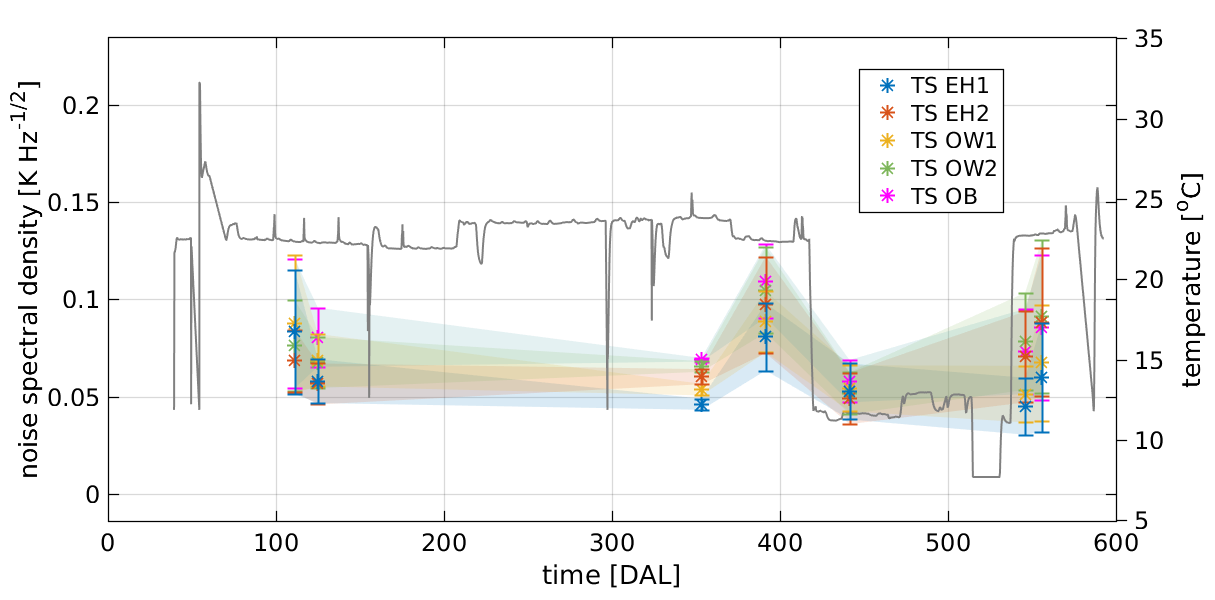}
\includegraphics[width=0.8\textwidth]{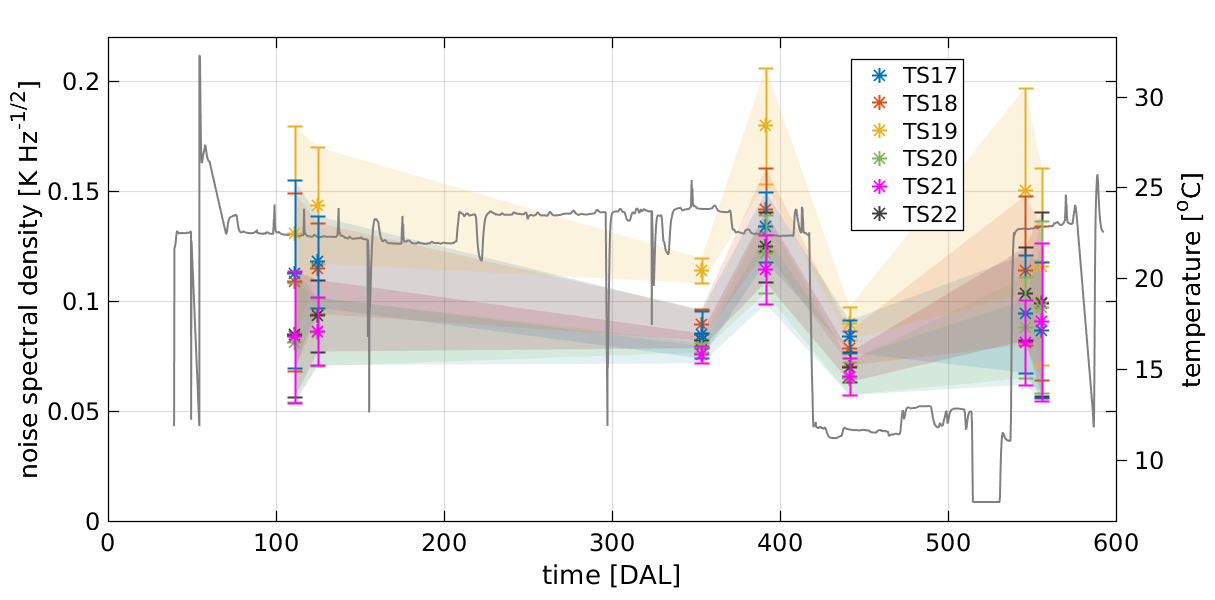}
\caption {Time evolution for the amplitude spectra of temperature fluctuations in the $\rm 10-30\, \mu Hz$ frequency range.
Only noise runs with several days of stable conditions. In grey we show the temperature profile of the mission 
for comparison. 
\emph{Top:} optical bench, optical window and electrode housing sensors. 
\emph{Bottom:} temperature sensors at the struts. 
\label{fig.k} }
\end{figure*}

So far our analysis focused on the 2017 February science run. 
However, in order to evaluate the non-stationarities in temperature fluctuations
we compared these results to the rest of science runs where the instrument
was kept unperturbed in its most sensitive configuration for several days.
Since we are purely interested in the temperature contribution, 
for each of these we compute the amplitude spectral density in the 
$\rm 10-30\, \mu Hz$ band. Results are shown in Figure~\ref{fig.k}. 
The amplitude of temperature fluctuations in the lowest bins of the LISA frequency band
is maintained in the $\rm 50-100\, mK/\sqrt{Hz}$ range for most of the runs
in the OB, OW and EH locations. 
As expected, the amplitude spectral density can increase up to $\rm 180\, mK/\sqrt{Hz}$
in some runs for the temperature in the struts as shown in the right panel of
Figure~\ref{fig.k}. 
The analysis shows therefore a considerable level of stationarity 
in the amplitude of temperature fluctuations in the $\rm 10-30\, \mu Hz$ band for the whole duration 
of the mission. The same conclusion could be drawn by comparing the power-law fits 
for each of the runs we have analysed. We refer to the interested reader to Appendix~\ref{sec.App_fits},
where we provide a table with the fits to power laws for each of these runs together with 
its dates of occurrence. 

Temperature fluctuations in these locations are hence described by the power 
law fits shown in the Figure~\ref{fig.TS_psd}. 
The rest of features appearing in the 
plot do not describe temperature fluctuations but are instead related to 
our temperature read-out. 
At higher frequencies the LISA Pathfinder temperature front end is limited by 
read-out noise, which is fundamentally dominated by the Wheastone bridge noise.
The in-flight measurements reached the design limit of 
 $\rm 10\,\mu K/\sqrt{Hz}$, a level that was also achieved during 
 on-ground testing~\citep{Lobo06a, Sanjuan07}.
This noise floor goes down to nearly 1\,mHz which was 
the LISA Pathfinder measuring bandwidth. 

The frequency regime spanning 0.2\, mHz$< f <$2\,mHz
is dominated by read-out noise arising from non-linearities in the temperature
diagnostics ADC~\citep{Sanjuan09c}. 
Although being studied and characterised during the design phase, 
this read-out noise source was not considered critical for the mission success 
since it was limited to frequencies below the LISA Pathfinder band.
This contribution can be modelled and subtracted to some extent~\citep{Sanjuan18}. 
Nevertheless the design of the future LISA temperature diagnostics subsystem 
will need to overcome this noise source that otherwise would affect 
the discrimination of temperature induced disturbances in the sub-milliHertz band. 
Since the ADC induced noise increases with wider excursion in the ADC range, a segment 
that explores less range will show a reduced impact of such a noise. 
The bottom plot in Figure~\ref{fig.TS_psd} confirms this by evaluating 
the amplitude spectral density of temperature fluctuations in a shorter segment with less temperature
drift.

Apart from the absolute temperature measurements, the temperature diagnostics front-end included 
some other channels to help disentangle noise sources in the actual temperature read-out. On one side, the so-called
\emph{reference} measurements were used to unambiguously determine the noise floor of the read-out.
These measurements were obtained by means of high stability resistors mounted on the same 
Wheatstone bridge used for the temperature sensing, allowing a direct measurement of the bridge electronics noise. 
On the other side, for some designated couple of sensors the electronics implemented a direct differential measurement 
by comparing directly them in the Wheatstone bridge, that is a direct hardware differential measurement. 
These were called \emph{differential} measurements. Both are shown in Figure~\ref{fig.TS_psd}.
The differential measurements are, in particular, a second cross-check to confirm the non-thermal origin of the 
excess noise observed in the mid-band. Indeed, due to its nature the differential measurements are
closer to zero and, hence, use less range of the ADC. As a consequence they are less exposed 
to ADC non-linearities induced noise, as it is shown in the plots.

\subsection{Thermal transfer functions}
\label{subsec.transfer_functions}

In the last section we evaluated the behaviour of temperature fluctuations and 
 quantified the noise spectra at different locations and its temporal evolution.
In the following we study the correlation between temperature fluctuations at
different locations. This is an important exercise to understand the thermal link between
the different subsystems on-board, a key aspect for an instrument like LPF and LISA.
We recall that the temperature diagnostic subsystem on-board included heaters 
in thermally sensitive locations such as the electrode housing, the optical windows and the struts. 
The objective of injecting temperature pulses at these locations was 
to study the instrument response to thermal disturbances in terms of forces exerted on the test mass 
or displacements measured by the optical read-out. 
The results of these experiments are of interest for the design of the future LISA mission 
and will be published elsewhere. However, as a side product, we can experimentally 
derive thermal transfer function between different locations, as we show below.  

To do so we use the different heat pulses that were injected in the 
different locations during the mission time line. 
The list of experiments is given in Table~\ref{tab.key_dates_experiments}
together with these, we will make use of the temperature \emph{bang-bang} control phase 
during the commissioning ---with temperature variations  $\sim 2^{\circ}$\,C--- produced 
by platform heaters. While the former will characterise point-to-point correlations between 
locations inside the experiment, the latter will tell us about the response of the diagnostic subsystem sensors 
 to external perturbations, a characterisation 
that can be of interest for the future LISA mission.

For each of these phases where an active thermal stimulus was present 
in the satellite, we derive the transfer functions 
comparing the Discrete Fourier Transform (DFT) 
of the temperature time series at two different locations. 
The result of this operation is a complex value 
that we show, expressed as magnitude and phase, by the dots in Figure~\ref{TFE_FIT}. 
The lines in the same plot correspond to the best fit models in frequency domain.
In Appendix~\ref{sec.App_transfer} we detail the estimation and 
modelling of the transfer functions and provide 
the parameters describing the models obtained during the analysis. 
It is important to keep in mind when interpreting these results 
that thermal injections were not intended for the 
purpose of our current study and, therefore, neither the amplitude nor the frequency 
of the applied signal are the optimal ones. 

\begin{figure}
\includegraphics[width=0.49\textwidth]{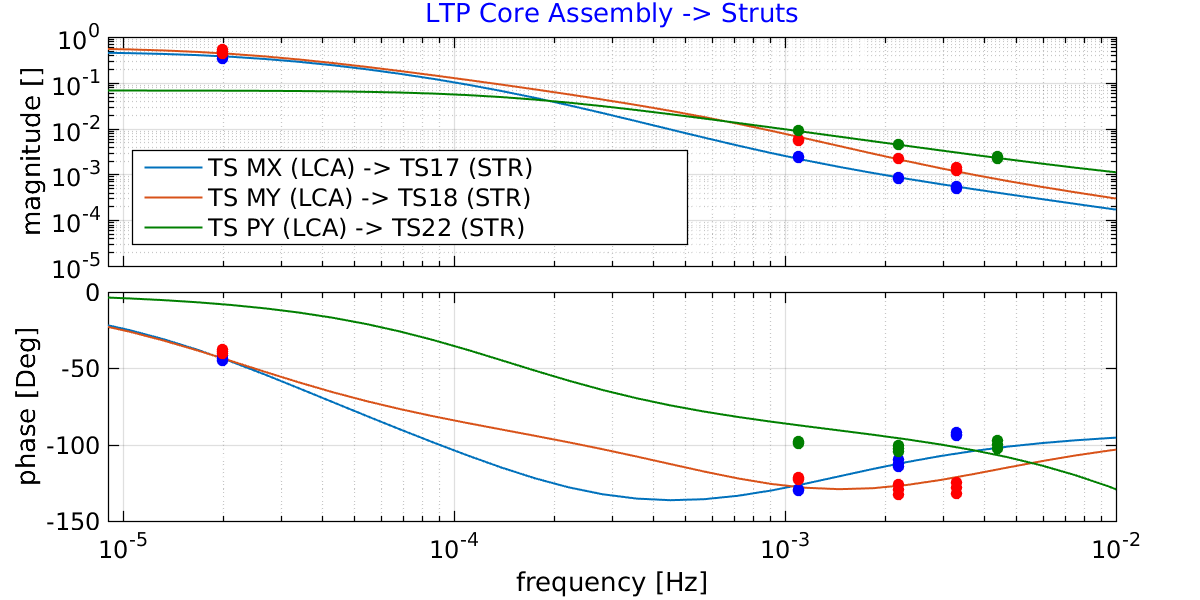}
\includegraphics[width=0.49\textwidth]{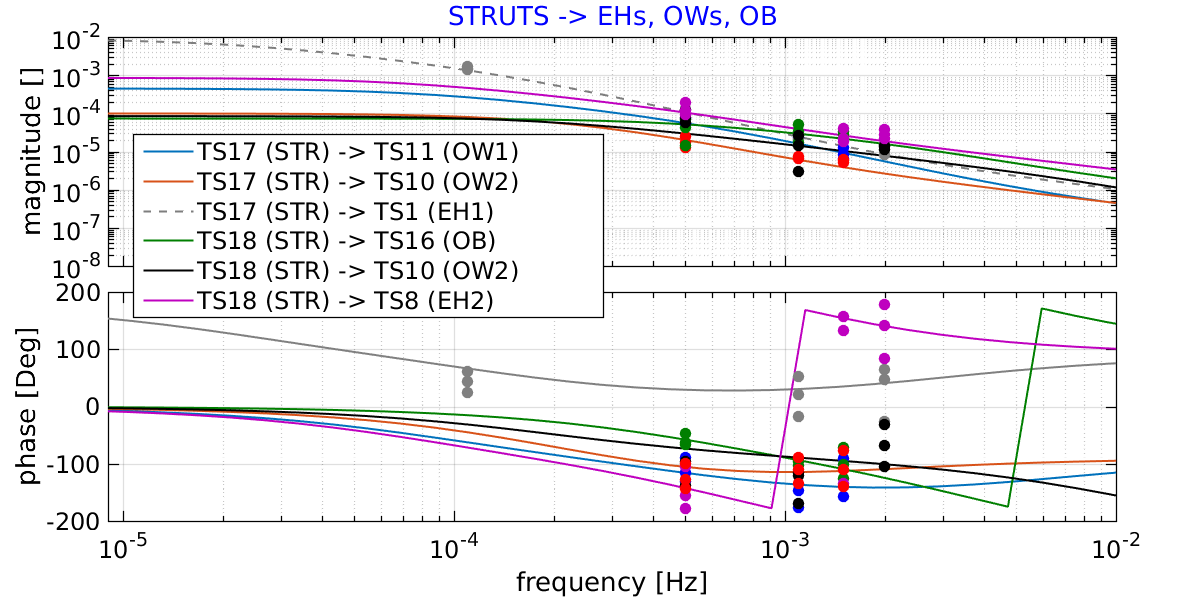}
\includegraphics[width=0.49\textwidth]{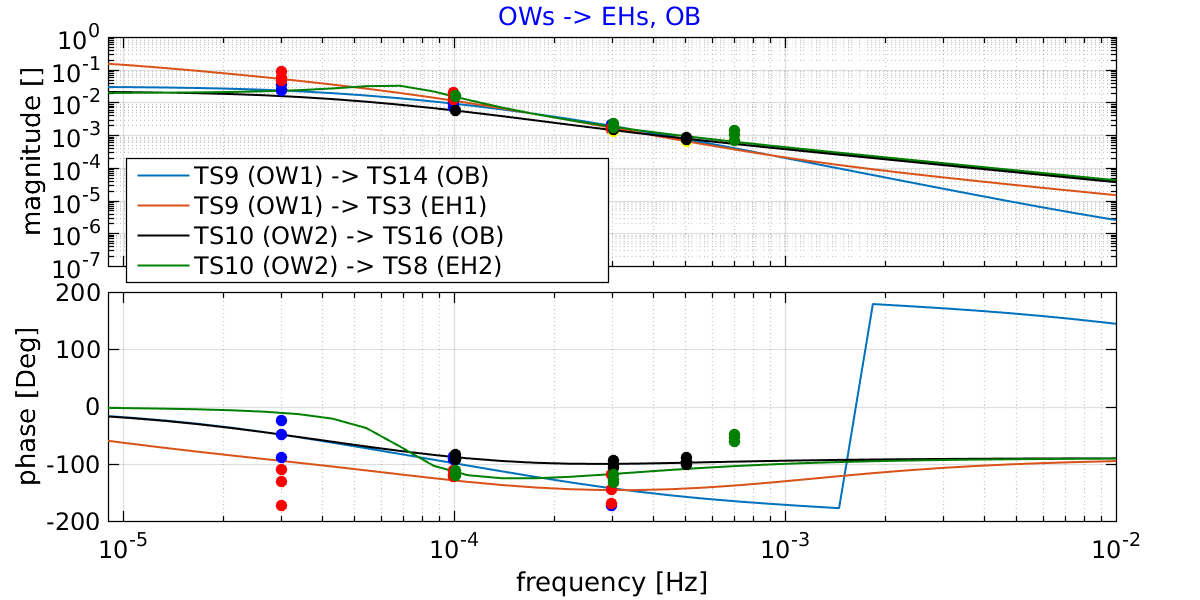}
\includegraphics[width=0.49\textwidth]{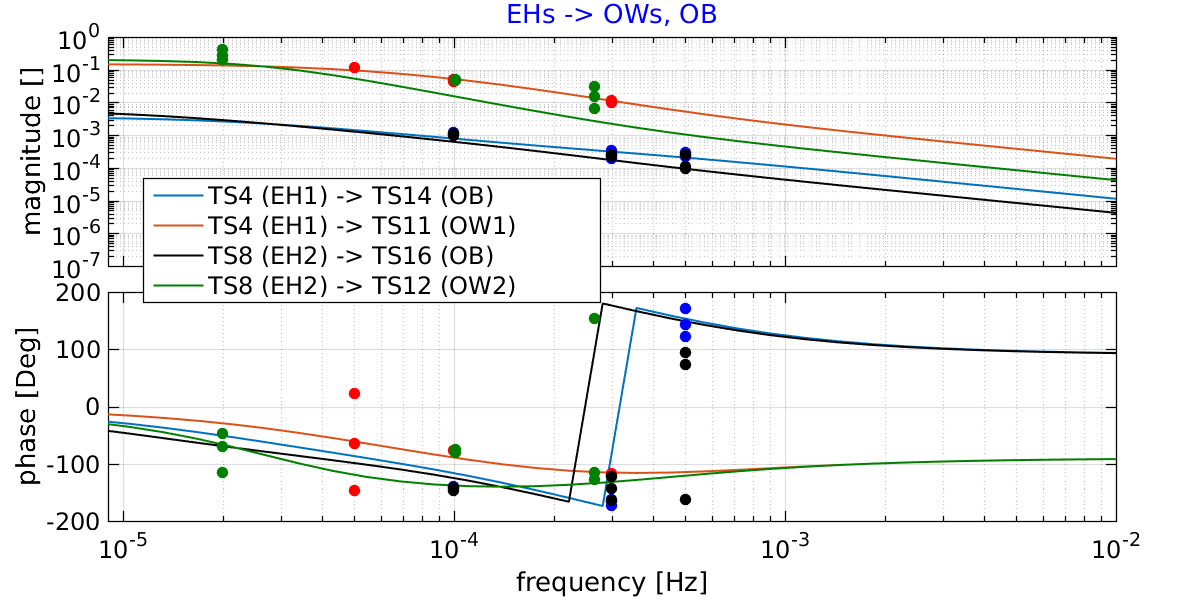}
\caption{Transfer functions between sensors in different locations. The points 
show these transfer functions at a certain frequency while the lines represent the fit to these functions. 
\label{TFE_FIT} }
\end{figure}

Figure~\ref{TFE_FIT} summarises our results. We show 
in the different panels the thermal transfer functions for each different 
location where the stimulus was applied. 
The top panel describes the effect of the thermal shield surrounding 
the main instrument onboard. 
We notice that the attenuation of temperature fluctuations from the external environment 
to the LISA Pathfinder instrument is better than $10^{-2}$ for fluctuations 
with frequencies above 1\,mHz. Temperature fluctuations are then 
further attenuated on their way to the inner part of the instrument 
when crossing the struts holding the instrument inside the thermal shield. Indeed, the second
panel shows the temperature suppression factor from the different struts to the rest of the locations
in the inner core of the instrument. Any temperature fluctuation moving through this path
in milliHertz equivalent time scales is attenuated by $5\cdot 10^{-5}$.
In this case, the CFRP (Carbon Fiber Reinforced Polymer) struts and the massive OB 
Zerodur structure are acting as equivalent thermal low pass filters. 

The last two panels describe the impact of thermal experiments ---as measured by surrounding sensors--- 
in the inner core of the instrument, i.e. the OW and the EH, respectively. 
A first point to take into account in understanding these figures is that, while the experiments in the OW reached temperature increases of $\sim 2^{\circ}$\,C, the temperature modulations in the EH ---being this a much more thermal sensitive locations--- 
were instead in the milliKelvin range. Consequently, the temperature increases due to the electrode 
housing experiments in the surrounding sensors are not so clearly measured and, thus, these transfer functions
are measured with lower precision.
Even though, thanks to the high precision of the temperature 
sensors we are able to estimate a $10^{-4}$ attenuation  in the LPF measuring band
for temperature fluctuations being transferred from the EH to the OB. 
The time scales characterising the thermal path for these fluctuations is $\sim 2$ days 
which makes easier to distinguish from other decorrelated temperature fluctuations.

Although not shown in the figures, we can get an estimate of the thermal transfer function at even lower frequencies. 
When taking into account the complete temperature series we observe a year modulation of the spacecraft solar array temperature of 
$\sim 3.5^{\circ}$\,C, which is proportional to the variation of the spacecraft solid angle with respect to the Sun throughout this period. 
The same modulation can be traced to the EH with an amplitude of $\sim 0.35^{\circ}$\,C, from where we derive a factor 10 attenuation to external temperature fluctuations in the frequency $3\cdot 10^{-8}$ Hz, that is to say a year period.

\section{Conclusions and implications for LISA}
\label{sec.conclusions}

The temperature diagnostics was a key subsystem of the LISA Pathfinder mission and was 
designed to disentangle the contribution of temperature fluctuations 
from the main metrology and force measurements. 
The temperature diagnostics subsystem consisted of 24 thermistors 
attached to sensitive locations and 16 heaters to produce
controlled inputs to calibrate the experiment response. 
The subsystem operated 
throughout the full mission duration, from the commissioning phase 
on 2016 January to the mission passivation on 2017 July. 
The sensors distributed across the LISA Pathfinder instrument 
allowed a precise characterisation of the temperature variations throughout the mission
and helped identify noise disturbances in the otherwise extremely stable 
environment in the satellite. 
We have detailed the reasons ---either intended or accidental--- of temperature 
variation in the mission timeline. 

Looking towards LISA, the most evident lesson learnt from the LISA Pathfinder
operations refers to the station-keeping manoeuvres. These were mandatory to keep 
the satellite in the Lissajous orbit in L1. However, the low thrust available in the 
propulsion system forced a several hours manoeuvre that produced an overall 
$\rm \sim 100\,mK$  temperature increase. Originally, this operation was repeated 
each weekend which, after the long thermal transient, left a few days for 
science runs in stable conditions. This periodic station-keeping 
is ruled out for LISA given that it would seriously impact its performance in the low frequency regime. 
Each LISA spacecraft will be injected in its individual orbit avoiding
the need of periodic corrections. Other temperature perturbations,
like the thermal experiments that were repeated frequently in a technology
demonstrator as LPF will be kept to the commissioning phase 
for LISA. 

In terms of performance, the diagnostic subsystem 
achieved the required performance of $\rm 10\,\mu K/\sqrt{Hz}$  
in the mission band, $\rm 1\, mHz< f < 30 \,mHz$,
showing only a slight deviation in the lowest frequency bin.
The latter is due to a coupling of the temperature drift on-board with 
non-linearities in the ADC. The effect was already 
known to effect the sub-milliHertz band during the design phase 
but the extensive operation period in a extremely quiet environment 
has allowed a precise characterisation. 
This will allow an improved 
design overcoming this read-out noise contribution for 
the future LISA temperature diagnostics subsystem.  

The low frequency band below the 100 $\mu Hz$
is dominated by temperature fluctuations that, in this band,
exceed the noise contribution from the ADC non-linearities. 
Thanks to the extensive data set we have been 
able to determine a noise level of
$\rm 50-100\,  mK/\sqrt{Hz}$  in the lowest bins of the LISA frequency band, 
$\rm 10-30\,  \mu Hz$, for those locations in the inner core of the experiment. 
This noise level was maintained during the different science runs throughout the mission,
which provides an important insight of the stationarity of the temperature 
fluctuations in the very low frequency domain during flight operations. 

We have also determined a $f^{-3.60 \pm 0.04}$ power law for the temperature fluctuations
in the electrode housing dominating the lowest frequency bins, below the $\rm 100 \,\mu Hz$. 
The cause of these fluctuations must be sought in the electronic units surrounding the
main instrument in the satellite. We have also determined fluctuations outside the 
thermal shield to be characterised by a $f^{-1.34 \pm 0.05}$ power law. 
The characterisation of these low frequency temperature fluctuations is 
relevant for LISA since this corresponds to the lowest frequency bins of the mission, 
where temperature variations are expected to provide a significant limit
 to the instrument's performance. 

These figures are however not directly applicable to LISA,
being the mission still in its definition phase, but
they are an important asset since they serve as an anchoring point for thermal design. 
In the same line of paving the way for the thermal design of the future LISA 
mission, we took advantage of the different thermal experiments in the mission 
time line to determine the thermal transfer function between locations, thereby
deriving the attenuation factors due to the different shielding layers
that can be used as guidance for the future LISA design.

\section*{Acknowledgments}

This work has been made possible by the LISA Pathfinder mission, 
which is part of the space-science program of the European Space Agency.
The French contribution has been supported by CNES (Accord Specific de 
projet CNES 1316634/CNRS 103747), the CNRS, 
the Observatoire de Paris and the University Paris-Diderot. 
E. P. and H. I. would also like to acknowledge the financial support of the UnivEarthS Labex program at Sorbonne Paris Cité (ANR-10-LABX-0023 and ANR-11- IDEX-0005-02).
The Albert-Einstein-Institut acknowledges the 
support of the German Space Agency, DLR. 
The work is supported by the Federal Ministry for Economic Affairs and Energy based on a resolution of the German Bundestag (FKZ 50OQ0501 and FKZ 50OQ1601).
The Italian contribution has been supported by 
Agenzia Spaziale Italiana and Instituto Nazionale di Fisica Nucleare. 
The Spanish contribution has been supported by Contracts No. AYA2010-15709 (MICINN), 
No. ESP2013-47637-P, and No. ESP2015-67234-P (MINECO). 
M. N. acknowledges support from Fundaci\'on General 
CSIC (Programa ComFuturo). F. R. acknowledges support from a Formaci\'on
de Personal Investigador (MINECO) contract.
The Swiss contribution acknowledges the support of 
the Swiss Space Office (SSO) via the PRODEX Programme 
of ESA. L. F. acknowledges the support of the Swiss National
Science Foundation.
The UK groups wish to acknowledge support from the
United Kingdom Space Agency (UKSA), the University of Glasgow, 
the University of Birmingham, Imperial College, 
and the Scottish Universities Physics Alliance (SUPA).
J.I.T. and J.S. acknowledge the support of the U.S. 
National Aeronautics and Space Administration (NASA).




\bibliographystyle{mnras}



\appendix

\section{Low frequency temperature power law fits}
\label{sec.App_fits}

\begin{table*}
   \resizebox{0.9\textwidth}{!}{
   \begin{tabular}{cccccccccc}
   \hline
          {Location} & {Parameters} & {Run \#1} & {Run \#2} & {Run \#3} & {Run \#4} & {Run \#5} & {Run \#6} & {Run \#7}\\ 
       \hline
         {EH1} & {$k$} & {-3.67 $\pm$ 0.03} & {-3.71 $\pm$ 0.02} & {-3.46 $\pm$ 0.05} & {-3.75 $\pm$ 0.02} & {-3.60 $\pm$ 0.04} & {-3.89 $\pm$ 0.05} & {-3.47 $\pm$ 0.06}\\
         {} & {$b$} & {(9 $\pm$ 3)$\cdot 10^{-17}$} & {(9 $\pm$ 2)$\cdot 10^{-17}$} & {(5 $\pm$ 3)$\cdot 10^{-16}$} & {(4.4 $\pm$ 0.7)$\cdot 10^{-17}$} & {(2.0 $\pm$ 0.8)$\cdot 10^{-16}$} & {(1.2 $\pm$ 0.5)$\cdot 10^{-17}$} & {(5 $\pm$ 3)$\cdot 10^{-16}$}\\              
         {EH2} & {$k$} & {-3.75 $\pm$ 0.05} & {-3.79 $\pm$ 0.03} & {-3.56 $\pm$ 0.06} & {-3.44 $\pm$ 0.03} & {-3.52 $\pm$ 0.04} & {-3.83 $\pm$ 0.08} & {-3.45 $\pm$ 0.02}\\
         {} & {$b$} & {(6 $\pm$ 3)$\cdot 10^{-17}$} & {(5 $\pm$ 1)$\cdot 10^{-17}$} & {(3 $\pm$ 2)$\cdot 10^{-16}$} & {(1.2 $\pm$ 0.4)$\cdot 10^{-15}$} & {(4 $\pm$ 1)$\cdot 10^{-16}$} & {(2 $\pm$ 1)$\cdot 10^{-17}$} & {(6 $\pm$ 1)$\cdot 10^{-16}$}\\                     
         {OB} & {$k$} & {-3.45 $\pm$ 0.09} & {-3.49 $\pm$ 0.05} & {-3.49 $\pm$ 0.05} & {-3.40 $\pm$ 0.04} & {-3.35 $\pm$ 0.05} & {-3.66 $\pm$ 0.06} & {-3.25 $\pm$ 0.06}\\
         {} & {$b$} & {9 $\pm$ 8)$\cdot 10^{-16}$} & {(1.0 $\pm$ 0.5)$\cdot 10^{-15}$} & {(9 $\pm$ 5)$\cdot 10^{-16}$} & {(1.9 $\pm$ 0.8)$\cdot 10^{-15}$} & {(2 $\pm$ 1)$\cdot 10^{-15}$} & {(1.3 $\pm$ 0.7)$\cdot 10^{-16}$} & {(5 $\pm$ 3)$\cdot 10^{-15}$}\\                   
         {OW1} & {$k$} & {-3.38 $\pm$ 0.04} & {-3.43 $\pm$ 0.03} & {-3.31 $\pm$ 0.07} & {-3.47 $\pm$ 0.03} & {-3.49 $\pm$ 0.04} & {-3.75 $\pm$ 0.07} & {-3.34 $\pm$ 0.04}\\
         {} & {$b$} & {(1.6 $\pm$ 0.6)$\cdot 10^{-15}$} & {(1.4 $\pm$ 0.34)$\cdot 10^{-15}$} & {(3 $\pm$ 2)$\cdot 10^{-15}$} & {(8 $\pm$ 2)$\cdot 10^{-16}$} & {(6 $\pm$ 3)$\cdot 10^{-16}$} & {(5 $\pm$ 3)$\cdot 10^{-17}$} & {(1.7 $\pm$ 0.6)$\cdot 10^{-15}$}\\                  
         {OW2} & {$k$} & {-3.70 $\pm$ 0.08} & {-3.50 $\pm$ 0.05} & {-3.34 $\pm$ 0.04} & {-3.20 $\pm$ 0.03} & {-3.21 $\pm$ 0.09} & {-3.80 $\pm$ 0.07} & {-3.29 $\pm$ 0.04}\\
         {} & {$b$} & {(1.0 $\pm$ 0.7)$\cdot 10^{-16}$} & {(8 $\pm$ 4)$\cdot 10^{-16}$} & {(3 $\pm$ 1)$\cdot 10^{-15}$} & {(1.2 $\pm$ 0.4)$\cdot 10^{-14}$} & {(9 $\pm$ 7)$\cdot 10^{-15}$} & {(3 $\pm$ 2)$\cdot 10^{-17}$} & {(3 $\pm$ 1)$\cdot 10^{-15}$}\\                              
         {TS17} & {$k$} & {-2.74 $\pm$ 0.07} & {-2.78 $\pm$ 0.04} & {-2.65 $\pm$ 0.05} & {-2.84 $\pm$ 0.03} & {-2.44 $\pm$ 0.06} & {-3.03 $\pm$ 0.09} & {-2.56 $\pm$ 0.06}\\       
         {} & {$b$} & {(9 $\pm$ 6)$\cdot 10^{-13}$} & {(9 $\pm$ 4)$\cdot 10^{-13}$} & {(2 $\pm$ 1)$\cdot 10^{-12}$} & {(5 $\pm$ 2)$\cdot 10^{-13}$} & {(2 $\pm$ 1)$\cdot 10^{-11}$} & {(6 $\pm$ 5)$\cdot 10^{-14}$} & {(3 $\pm$ 2)$\cdot 10^{-12}$}\\                      
         {TS18} & {$k$} & {-2.73 $\pm$ 0.09} & {-2.81 $\pm$ 0.04} & {-2.73 $\pm$ 0.07} & {-2.85 $\pm$ 0.05} & {-2.54 $\pm$ 0.08} & {-2.92 $\pm$ 0.08} & {-2.42 $\pm$ 0.09}\\       
         {} & {$b$} & {(1.0 $\pm$ 0.9)$\cdot 10^{-12}$} & {(7 $\pm$ 3)$\cdot 10^{-13}$} & {(1.1 $\pm$ 0.8)$\cdot 10^{-12}$} & {(5 $\pm$ 2)$\cdot 10^{-13}$} & {(7 $\pm$ 2)$\cdot 10^{-12}$} & {(2 $\pm$ 1)$\cdot 10^{-13}$} & {(2 $\pm$ 1)$\cdot 10^{-11}$}\\                            
         {TS19} & {$k$} & {-2.70 $\pm$ 0.07} & {-2.83 $\pm$ 0.05} & {-2.75 $\pm$ 0.06} & {-2.76 $\pm$ 0.07} & {-2.18 $\pm$ 0.06} & {-3.10 $\pm$ 0.08} & {-2.52 $\pm$ 0.08}\\       
         {} & {$b$} & {(2 $\pm$ 1)$\cdot 10^{-12}$} & {(7 $\pm$ 3)$\cdot 10^{-13}$} & {(1.1 $\pm$ 0.6)$\cdot 10^{-12}$} & {(1.3 $\pm$ 0.9)$\cdot 10^{-12}$} & {(3 $\pm$ 2)$\cdot 10^{-10}$} & {(4 $\pm$ 3)$\cdot 10^{-14}$} & {(7 $\pm$ 5)$\cdot 10^{-12}$}\\                        
         {TS20} & {$k$} & {-2.83 $\pm$ 0.08} & {-2.98 $\pm$ 0.06} & {-2.88 $\pm$ 0.05} & {-2.89 $\pm$ 0.06} & {-2.53 $\pm$ 0.05} & {-2.93 $\pm$ 0.06} & {-2.47 $\pm$ 0.06}\\       
         {} & {$b$} & {(4 $\pm$ 3)$\cdot 10^{-13}$} & {(1.3 $\pm$ 0.7)$\cdot 10^{-13}$} & {(3 $\pm$ 1)$\cdot 10^{-13}$} & {(3 $\pm$ 2)$\cdot 10^{-13}$} & {(7 $\pm$ 4)$\cdot 10^{-12}$} & {(2 $\pm$ 1)$\cdot 10^{-13}$} & {(9 $\pm$ 5)$\cdot 10^{-12}$}\\                           
         {TS21} & {$k$} & {-2.82 $\pm$ 0.07} & {-2.93 $\pm$ 0.02} & {-2.77 $\pm$ 0.06} & {-2.96 $\pm$ 0.02} & {-2.64 $\pm$ 0.05} & {-2.94 $\pm$ 0.05} & {-2.62 $\pm$ 0.07}\\       
         {} & {$b$} & {(4 $\pm$ 3)$\cdot 10^{-13}$} & {(2.1 $\pm$ 0.4)$\cdot 10^{-13}$} & {(7 $\pm$ 4)$\cdot 10^{-13}$} & {(1.5 $\pm$ 0.3)$\cdot 10^{-13}$} & {(2 $\pm$ 1)$\cdot 10^{-12}$} & {(1.7 $\pm$ 0.7)$\cdot 10^{-13}$} & {(2 $\pm$ 1)$\cdot 10^{-12}$}\\
         {TS22} & {$k$} & {-2.82 $\pm$ 0.07} & {-2.94 $\pm$ 0.04} & {-2.80 $\pm$ 0.04} & {-2.88 $\pm$ 0.05} & {-2.50 $\pm$ 0.05} & {-2.87 $\pm$ 0.06} & {-2.55 $\pm$ 0.08}\\       
         {} & {$b$} & {(4 $\pm$ 3)$\cdot 10^{-13}$} & {(1.9 $\pm$ 0.7)$\cdot 10^{-13}$} & {(5 $\pm$ 2)$\cdot 10^{-13}$} & {(3 $\pm$ 2)$\cdot 10^{-13}$} & {(1.0 $\pm$ 0.4)$\cdot 10^{-11}$} & {(4 $\pm$ 2)$\cdot 10^{-13}$} & {(4 $\pm$ 3)$\cdot 10^{-12}$}\\
   \end{tabular}}
   \caption{Parameters for the power law fit of the temperature fluctuations amplitude spectral density at low frequencies. The model is given in Equation~(\ref{eq.powerlaw}) and the different runs described in the text. }
          \label{tab.fit_PSD} 
\end{table*}

We compute the amplitude spectral density 
by means of the Welch averaged periodogram. 
We use segments of 400,000\,s and apply 
Blackman-Harris window to prevent from  spectral leakage. 
After subtracting the lowest four frequency bins, we perform 
a power law fit at the (remaining) four lowest frequency bins. 

The expression we use for the fit is given by
\begin{equation}
S^{1/2}(f) = b\cdot (2 \pi f)^{k}
\label{eq.powerlaw}
\end{equation}
 where $S^{1/2}(f)$ is the amplitude spectral density and $b$ and $k$ are the parameters of the fit.
In the locations where there is more than one sensor we 
use an average of sensors when considering the power law fit,
these are the electrode housing (EH1 and EH2) with four sensors in each of them;
the optical windows (OW1 and OW2) with three sensors in each;
and the optical bench (OB) with four sensors, one in each corner. 
The six remaining (TS17-TS22) correspond to sensors attached to 
different struts. We recall that only six out of the eight struts had a pair heater/sensor attached.

In Table~\ref{tab.fit_PSD} we provide the results of the fits for each of the noise runs during the
LTP operations phase. These were periods were the instruments 
was configured in its optimal sensitivity configuration and left unperturbed
during days and even weeks in some cases. 
For the sake of completeness we list here below the periods were these
runs took place. 

\begin{itemize}
\item Run \#1 $\rightarrow$ from March 20 to March 26 (2016)
\item Run \#2 $\rightarrow$ from April 3 to April 16 (2016)
\item Run \#3 $\rightarrow$ from November 17 to November 26 (2016)
\item Run \#4 $\rightarrow$ from December 26 (2016) to January 13 (2017)
\item Run \#5 $\rightarrow$ from February 14 to February 27 (2017)
\item Run \#6 $\rightarrow$ from May 29 to June 5 (2017)
\item Run \#7 $\rightarrow$ from June 8 to June 17 (2017)
\end{itemize}

The results of the power-law fits in Table~\ref{tab.fit_PSD} support 
the ones previously showed in Figure~\ref{fig.k}.  
In that case, we provided the time evolution 
for the amplitude spectra of temperature 
fluctuations in the $\rm 10-30\, \mu Hz$ frequency range
for the same noise runs. 
We consider both, the power-law fit and the noise density in the low frequency regime,
providing a complementary view of the same phenomena, 
which can be of interest for the reader. 

\section{Thermal transfer functions}
\label{sec.App_transfer} 

\begin{table*}
   \begin{tabular}{ccccccc}
   \hline
   Origin   $\rightarrow$ End    & {$r_{1}$} & {$p_{1}$} & {$r_{2}$} & {$p_{2}$} & corner frequency ($\mu$Hz) & DC gain\\        
\hline   
       LCA4 $\rightarrow$ TS17 &   {$-8\cdot 10^{-5}$} & {$-9\cdot 10^{-4}$} & {$9\cdot 10^{-5}$} & {$-2\cdot 10^{-4}$} & 36 & {$3.6\cdot 10^{-1}$}\\
        LCA5 $\rightarrow$ TS18 & {$-7\cdot 10^{-5}$} & {$-4\cdot 10^{-3}$} & {$8\cdot 10^{-5}$} & {$-1\cdot 10^{-4}$} & 54 & {$7.8\cdot 10^{-1}$}\\
         LCA2 $\rightarrow$ TS22 & {$-3\cdot 10^{-4}$} & {$-3\cdot 10^{-1}$} & {$6\cdot 10^{-5}$} & {$-9\cdot 10^{-4}$} & 140 & {$6.6\cdot 10^{-2}$}\\
         TS17 $\rightarrow$ TS10 & {$-2\cdot 10^{-7}$} & {$-2\cdot 10^{-3}$} & {$2\cdot 10^{-7}$} & {$-1\cdot 10^{-3}$} & 122 & $1.0\cdot 10^{-4}$\\
         TS17 $\rightarrow$ TS11 & {$-2\cdot 10^{-7}$} & {$-4\cdot 10^{-3}$} & {$2\cdot 10^{-7}$} & {$-5\cdot 10^{-4}$} & 350 & $1.2\cdot 10^{-4}$\\
         TS17 $\rightarrow$ TS1 & {$1\cdot 10^{-6}$} & {$-9\cdot 10^{-4}$} & {$-1\cdot 10^{-6}$} & {$-1\cdot 10^{-4}$} & 26 & $8.9\cdot 10^{-3}$\\
         TS18 $\rightarrow$ TS16 & {$-4\cdot 10^{-7}$} & {$-2\cdot 10^{-2}$} & {$3\cdot 10^{-7}$} & {$-3\cdot 10^{-3}$} & 1413 & $9.6\cdot 10^{-5}$\\
         TS18 $\rightarrow$ TS10 & {$-1\cdot 10^{-7}$} & {$-6\cdot 10^{-2}$} & {$1\cdot 10^{-7}$} & {$-1\cdot 10^{-3}$} & 666 & $9.8\cdot 10^{-4}$\\
        TS18 $\rightarrow$ TS8 &  {$-7\cdot 10^{-7}$} & {$-4\cdot 10^{-3}$} & {$5\cdot 10^{-7}$} & {$-5\cdot 10^{-4}$} & 122 & $8.3\cdot 10^{-4}$\\
        TS9 $\rightarrow$ TS14 &  {$-9\cdot 10^{-6}$} & {$-1\cdot 10^{-3}$} & {$9\cdot 10^{-6}$} & {$-2\cdot 10^{-4}$} & 38 & $3.6\cdot 10^{-2}$\\
         TS9 $\rightarrow$ TS3 & {$-1\cdot 10^{-5}$} & {$-6\cdot 10^{-4}$} & {$1\cdot 10^{-5}$} & {$-4\cdot 10^{-5}$} & 13 & $2.3\cdot 10^{-1}$\\
        TS10 $\rightarrow$ TS16 &  {$-3\cdot 10^{-6}$} & {$-8\cdot 10^{-4}$} & {$5\cdot 10^{-6}$} & {$-2\cdot 10^{-4}$} & 34 & $2.1\cdot 10^{-2}$\\
        TS10 $\rightarrow$ TS8 &  {$3\cdot 10^{-4}$} & {$-2$} & {$4\cdot 10^{-6}$} & {$-4\cdot 10^{-5}$} & 769 & $1.0\cdot 10^{-1}$\\
        TS4 $\rightarrow$ TS14 &  {$-1\cdot 10^{-6}$} & {$-2\cdot 10^{-3}$} & {$6\cdot 10^{-7}$} & {$-1\cdot 10^{-4}$} & 38 & $5.5\cdot 10^{-3}$\\
        TS4 $\rightarrow$ TS11 &  {$-3\cdot 10^{-5}$} & {$-2\cdot 10^{-3}$} & {$3\cdot 10^{-5}$} & {$-2\cdot 10^{-4}$} & 54 & $1.4\cdot 10^{-1}$\\
       TS8 $\rightarrow$ TS16 &  {$-7\cdot 10^{-7}$} & {$-1\cdot 10^{-3}$} & {$5\cdot 10^{-7}$} & {$-7\cdot 10^{-5}$} & 23 & $6.4\cdot 10^{-3}$\\
        TS8 $\rightarrow$ TS12 &  {$-2\cdot 10^{-5}$} & {$-1\cdot 10^{-3}$} & {$4\cdot 10^{-5}$} & {$-9\cdot 10^{-5}$} & 26 & $4.2\cdot 10^{-1}$\\
   \end{tabular}
   \caption{Thermal transfer functions parameters between different locations. The model is given in Equation~(\ref{eq.model_tfe}).}
          \label{tab.fit_TFE} 
\end{table*}

The thermal transfer function between an origin location A and a final
location B at a given frequency $f $ is experimentally computed as 

\begin{equation}
H_{A \rightarrow B} (f)=  \frac{\widetilde T_B (f )}{ \widetilde T_A (f )}
\label{eq.tfe}
\end{equation}

where $\widetilde T_A (f)$ and $\widetilde T_B (f)$ are the Discrete Fourier Transform (DFT)
of the temperatures at the location A and B at a given frequency $f$.
Transfer functions are thus only computed between locations linked with pairs of sensors. 
In order to be representative of the heat flow between locations, these are
just estimated when a heat input is active at the origin location, in
which case we can make use of the temperature sensor close to the 
heater as representative of the heat injection.  
As previously commented, the transfer functions from the outside of the thermal shield to the 
inside are derived using the characterisation provided by the \emph{bang-bang} controller 
during the commissioning. These was a homogeneous temperature modulation of all the spacecraft from where 
we can extract thermal transfer functions from a  temperature sensor outside the thermal shield to its 
closest counterpart inside, typically attached to a strut.

Once we have derived the points that experimentally define the transfer function, 
we can fit them to a continuous model given by 
\begin{equation}
H_{A \rightarrow B}(s)=\frac{r_{1}}{s-p_{1}}+\frac{r_{2}}{s-p_{2}}
\label{eq.model_tfe}
\end{equation}
where $H_{A \rightarrow B}(s)$ is the Laplace transform of 
the differential equation that describes the heat flow between points A and B. 
In our case, a second order transfer function model 
described by residuals, $r_{1}$ and $r_{2}$, and poles, $p_{1}$ and $p_{2}$.
This expression corresponds to a differential equation 
that can be understood as an approximation to second order of the heat flow
equation that describes the heat flow from the origin to the final location. 
The fit is done using the vector fit algorithm~\citep{Gustavsen99}
implemented in the LTPDA toolbox~\citep{Hewitson09}. 
In Table~\ref{tab.fit_TFE} we report the values obtained for these fits
and in Figure~\ref{TFE_FIT} we show a representative set of these types of transfer functions.

\medskip  
 $^{a}$  European Space Technology Centre, European Space Agency, 
Keplerlaan 1, 2200 AG Noordwijk, The Netherlands \\
 $^{b}$  Albert-Einstein-Institut, Max-Planck-Institut f\"ur Gravitationsphysik und Leibniz Universit\"at Hannover,
Callinstra{\ss}e 38, 30167 Hannover, Germany \\
 $^{c}$  APC, Univ Paris Diderot, CNRS/IN2P3, CEA/lrfu, Obs de Paris, Sorbonne Paris Cit\'e, France \\
 $^{d}$  Department of Industrial Engineering, University of Trento, via Sommarive 9, 38123 Trento, 
and Trento Institute for Fundamental Physics and Application / INFN \\
 $^{e}$  Dipartimento di Fisica, Universit\`a di Trento and Trento Institute for 
Fundamental Physics and Application / INFN, 38123 Povo, Trento, Italy \\
 $^{f}$  Istituto di Fotonica e Nanotecnologie, CNR-Fondazione Bruno Kessler, I-38123 Povo, Trento, Italy \\
 $^{g}$  DISPEA, Universit\`a di Urbino ``Carlo Bo'', Via S. Chiara, 27 61029 Urbino/INFN, Italy \\
 $^{h}$  The School of Physics and Astronomy, University of
Birmingham, Birmingham, UK \\
 $^{i}$  European Space Astronomy Centre, European Space Agency, Villanueva de la
Ca\~{n}ada, 28692 Madrid, Spain \\
 $^{j}$  Institut f\"ur Geophysik, ETH Z\"urich, Sonneggstrasse 5, CH-8092, Z\"urich, Switzerland \\
 $^{k}$  The UK Astronomy Technology Centre, Royal Observatory, Edinburgh, Blackford Hill, Edinburgh, EH9 3HJ, UK \\
 $^{l}$  Institut de Ci\`encies de l'Espai (CSIC-IEEC), Campus UAB, Carrer de Can Magrans s/n, 08193 Cerdanyola del Vall\`es, Spain \\
 $^{m}$  European Space Operations Centre, European Space Agency, 64293 Darmstadt, Germany \\
 $^{n}$  High Energy Physics Group, Physics Department, Imperial College London, Blackett Laboratory, Prince Consort Road, London, SW7 2BW, UK \\
 $^{o}$  Department of Mechanical and Aerospace Engineering, MAE-A, P.O. Box 116250, University of Florida, Gainesville, Florida 32611, USA \\
 $^{p}$  Physik Institut, 
Universit\"at Z\"urich, Winterthurerstrasse 190, CH-8057 Z\"urich, Switzerland \\
 $^{q}$  SUPA, Institute for Gravitational Research, School of Physics and Astronomy, University of Glasgow, Glasgow, G12 8QQ, UK \\
 $^{r}$  Department d'Enginyeria Electr\`onica, Universitat Polit\`ecnica de Catalunya,  08034 Barcelona, Spain \\
 $^{s}$  Deutsches Zentrum f\"ur Luft- und Raumfahrt, Robert-Hooke-Str. 7, 28359 Bremen, Germany \\
 $^{t}$  Gravitational Astrophysics Lab, NASA Goddard Space Flight Center, 8800 Greenbelt Road, Greenbelt, MD 20771 USA \\
 $^{u}$  Airbus Defence and Space, Gunnels Wood Road, Stevenage, Hertfordshire, SG1 2AS, United Kingdom \\


\bsp	
\label{lastpage}
\end{document}